# The unitary theory of the electrical powers

**Gheorghe Mihai**

*University of Craiova, Faculty of Electrotechnics, Bdul Decebal 5, Romania*

*E-mail:* gmihai@elth.ucv.ro

Abstract: General physics approach is applied to analysis of power components in electrical systems under sinusoidal and non-sinusoidal conditions. Physical essence of active, reactive and distorting powers are determinate. It is shown that the all three powers are the different aspects of the same physical phenomenon: mechanical action per time of Coulomb forces or inertial forces.

KEYWORDS: Electrical Powers

# Contents

## 1. Introduction

In 1927, C. Budeanu has introduced for the first time the reactive and distorting powers, not based by physical phenomenon, only based by definitions. Many studies have analyzed the physical significances, but the problem has not resolved. Thus, there are many definitions in literature [1], [2], [3] and different approach. In this work, we are starting with following physical fundamental concepts for resolving the electrical powers under sinusoidal and non-sinusoidal conditions:

A1) We can associate a mass to any energy form, based by Einstein formula: $E = mc_0^2$

A2) The inertial force represents the variation of the impulse under time: $F_{in} = \dfrac{dp}{dt}$

A3) The instantaneous power is the product between force and velocity, at a time moment.

## I. The physical theory of the electrical powers in Single-Phases electrical circuits in sinusoidal and non-sinusoidal periodic regime

### I.1. The electrical active power – *P*

Let us consider an electrical current $i(t)$ that pass throw a material conductor. The current density $\bar{J}(t)$ and the mean velocity of electrical charges $\bar{v}(t)$ at a moment time are in relation:

$$\overline{J}(t) = \rho\overline{v}(t) \tag{I.1}$$

The force exercised on the electrical density of the electrical field $\overline{E}(t)$ is:

$$\overline{F}(t) = \rho\overline{E}(t) \tag{I.2}$$

An instantaneous power is determined as:

$$p(t) = \overline{F}(t) \cdot \overline{v}(t) \tag{I.3}$$

By substituting $F(t)$ from (I.2) in (I.3) will obtain:

$$p(t) = \rho \cdot \overline{E}(t) \cdot \overline{v}(t) \tag{I.4}$$

Using (I.1) will obtain:

$$p(t) = \overline{J}(t) \cdot \overline{E}(t) \tag{I.5}$$

Relation (I.5) represents the power density in a point situated in $\Omega$ domain, covering of the electrical current. The instantaneous total power in the $\Omega$ domain is equal:

$$P_{tot}(t) = \int_{\Omega} p(t) dv \tag{I.6}$$

The instantaneous total power has obtained by substituting (I.5) and the elemental volume $dv = d\overline{S} d\overline{l}$ in (I.6):

$$P_{tot}(t) = \int_{1}^{2}\int_{S} \left(\overline{J}(t)d\overline{S}\right)\left(\overline{E}(t)d\overline{l}\right) = u(t) \cdot i(t) \tag{I.7}$$

The mean power averaged over time interval is:

$$P_{mean} = \frac{1}{t_2 - t_1}\int_{t_1}^{t_2} u(t)i(t)dt = \langle u(t), i_{//}(t) + i_{\perp}(t)\rangle = \|u(t)\| \cdot \|i_{//}(t)\| \tag{I.8}$$

Where $i_{//}(t)$ and $i_{\perp}(t)$ are the orthogonal components of the $i(t)$ in relation with $u(t)$ [5].
Under sinusoidal conditions:

$$\begin{cases} u(t) = U\sqrt{2}\sin\omega t \\ i(t) = I\sqrt{2}\sin(\omega t - \varphi) \end{cases} \tag{I.9}$$

Using (I.8) and (I.9) the mean power averaged over a period is equal:

$$P_{mean} = UI\cos\varphi \tag{I.10}$$

Using non-sinusoidal conditions:

$$\begin{cases} u(t) = U_0 + \sum_{k=1}^{\infty} U_k \sqrt{2} \sin(k\omega t + \alpha_k) \\ i(t) = I_0 + \sum_{k=1}^{\infty} I_k \sqrt{2} \sin(k\omega t + \alpha_k - \varphi_k) \end{cases} \quad (I.11)$$

The mean power over a period is equal:

$$P_{mean} = U_0 I_0 + \sum_{k=1}^{\infty} U_k I_k \cos\varphi_k \quad (I.12)$$

Analysing (I.10) and (I.12) the conclusion drawn is that the mean active power is the mean work done by the coulombian electrical forces over a period.

### I.2. The electrical reactive power - Q

The process in the circuit with **R** and **L** elements connected in series will analyzed. The instantaneous magnetically energy of the field is equal:

$$W_m(t) = \frac{Li^2(t)}{2} \quad (I.13)$$

The magnetic energy is a function of the velocity of the electrical charges, from the current density *i(t)*. The literature [4] show that the Poynting vector that depends to electrical charges velocity does not radiated energy. According (I.13), the magnetic energy stored in the inductor is non-radiate but is only variable in time.

According to A1) condition, the "field mass" - $m_{field}$ associated to (I.13) is moving with the speed of light. So:

$$m_{field}(t) = \frac{Li^2(t)}{2c_0^2} \quad (I.14)$$

The associated impulse - $\bar{p}_{field}$ to the "field mass" is equal:

$$\bar{p}_{field}(t) = m_{field}(t)\bar{c}_0 = \frac{Li^2(t)}{2c_0^2}\bar{c}_0 \quad (I.15)$$

According to A2), the rate of the impulse is equal to inertial force - $\bar{F}_{in}$:

$$\bar{F}_{in}(t) = \frac{d\bar{p}_{field}}{dt} = \bar{c}_0 \frac{dm_{field}(t)}{dt} \quad (I.16)$$

The instantaneous power corresponding to the displacement of the "field mass" is equal:

$$Q_{in}(t) = \bar{c}_0 \bar{F}_{in}(t) \quad (I.17)$$

The inertial force may be both positive and negative. We join the positive sign of the inertial force to the inductor loading with magnetically energy and the minus sign then the inductor unload. The necessary power for load and unload the inductor id deliberated by the electric generator. We consider the effective value of the inertial force both for loading and for unloading of the inductor:

$$F_{in.ef} = \sqrt{\frac{1}{T}\int_0^T F_{in}^2(t)dt} \qquad (I.18)$$

The reactive power is the multiplied between the speed of light and the effective inertial force:

$$Q = c_0 F_{in.ef} \qquad (I.19)$$

**I.2.1. The reactive power for an inductor**

Starting with the sinusoidal current:

$$i(t) = I\sqrt{2}\sin\omega t \qquad (I.20)$$

Then, the "field mass" stored into magnetic field of the inductor is:

$$m_{field}(t) = \frac{LI^2}{c_0^2}\sin^2\omega t \qquad (I.21)$$

In addition, the corresponding impulse is equal:

$$p_{field}(t) = \frac{LI^2}{c_0}\sin^2\omega t \qquad (I.22)$$

The inertial force at a moment generated by the "field mass" is:

$$F_{in}(t) = \frac{dp_{field}}{dt} = \frac{\omega LI^2}{c_0}\sin 2\omega t \qquad (I.23)$$

Let consider the interval (**0, T/4**) when is produced the energy loading of the inductor, and calculate the effective value of the inertial force:

$$F_{in.ef} = \sqrt{\frac{2}{T}\int_0^{T/4}\left(\frac{\omega LI^2}{c_0}\sin 2\omega t\right)^2 dt} = \frac{\omega LI^2}{2c_0} \qquad (I.24)$$

The reactive power necessary to load the inductor is equal:

$$Q_{load} = c_0 F_{in.ef} = \frac{\omega LI^2}{2} \qquad (I.25)$$

Similarly, the reactive power necessary to unload the inductor is equal:

$$Q_{unload} = c_0\sqrt{\frac{2}{T}\int_{T/4}^{T/2}\left(\frac{\omega LI^2}{c_0}\sin 2\omega t\right)^2 dt} = \frac{\omega LI^2}{2} \qquad (I.26)$$

The total reactive power is the sum between (I.25) and (I.26):

$$Q = \omega L I^2 \qquad (I.27)$$

Considering the electric parameters of the circuit, the above relation can write:

$$Q = UI \sin \varphi \qquad (I.28)$$

### I.2.2. The reactive power for a capacitor

Starting with the same electrical current (I.20), we consider a capacitor circuit.
The tension for the capacitor is equal:

$$u_C(t) = \frac{1}{C} \int i \, dt = -\frac{I\sqrt{2}}{\omega C} \cos \omega t \qquad (I.29)$$

The "field mass" stored into electric field of the capacitor is:

$$m_{field}(t) = \frac{C u_C^2(t)}{2 c_0^2} = \frac{I^2}{c_0^2 \omega^2 C} \cos^2 \omega t \qquad (I.30)$$

The inertial force at a moment generated by the "field mass" is:

$$F_{in}(t) = -\frac{I^2}{c_0 \omega C} \sin 2\omega t \qquad (I.31)$$

Let consider the interval (**0, T/4**) when is produced the energy loading of the capacitor, and calculate the effective value of the inertial force:

$$F_{in.ef} = \sqrt{\frac{2}{T} \int_0^{T/4} \left(-\frac{I^2}{c_0 \omega C} \sin 2\omega t\right)^2 dt} = \frac{I^2}{2 c_0 \omega C} \qquad (I.32)$$

The reactive power necessary to load the capacitor is equal:

$$Q_{load} = c_0 F_{in.ef} = \frac{I^2}{2\omega C} \qquad (I.33)$$

Similarly, the reactive power necessary to unload the inductor is equal:

$$Q_{unload} = c_0 \sqrt{\frac{2}{T} \int_{T/4}^{T/2} \left(-\frac{I^2}{c_0 \omega C} \sin 2\omega t\right)^2 dt} = \frac{I^2}{2\omega C} \qquad (I.34)$$

The total reactive power is the sum between (I.33) and (I.34):

$$Q = \frac{I^2}{\omega C} \qquad (I.35)$$

### I.2.3. The reactive power in LC circuit

Comparing (I.23) and (I.31) we observe that the inertial force produced into inductor has opposite sign compare that the capacitor.
The sum of (I.23) and (I.31) is equal:

$$F_{in.rez.}(t) = \frac{I^2}{c_0}\left(\omega L - \frac{1}{\omega C}\right)\sin 2\omega t \tag{I.36}$$

The reactive power is:

$$Q = 2c_0\sqrt{\frac{2}{T}\int_0^{T/4}\left[\frac{I^2}{c_0}\left(\omega L - \frac{1}{\omega C}\right)\sin 2\omega t\right]^2 dt} = I^2\left(\omega L - \frac{1}{\omega C}\right) \tag{I.37}$$

We can write the above relation as:

$$Q = UI\sin\varphi_1 - UI\sin\varphi_2 \tag{I.38}$$

### I.3. The electrical distorting power-D

Let consider an electrical circuit operated under non-sinusoidal conditions:

$$u(t) = U_0 + \sum_{k=1}^{\infty} U_k \sin(k\omega t + \alpha_k) \tag{I.39}$$

$$i(t) = I_0 + \sum_{k=1}^{\infty} I_k \sqrt{2}\sin(k\omega t + \alpha_k - \varphi_k) \tag{I.40}$$

According with C. Budeanu theory, the distorting power is equal:

$$D^2 = \sum_{\substack{m=1 \\ n\neq m}}^{\infty}\sum_{n=1}^{\infty}\left[(U_m I_n)^2 + (U_n I_m)^2\right] - \sum_{\substack{m=1 \\ n\neq m}}^{\infty}\sum_{n=1}^{\infty}\left[2U_m U_n I_m I_n \cos(\varphi_m - \varphi_n)\right] \tag{I.41}$$

Starting with the following relations:

$$\cos(\varphi_m - \varphi_n) = \cos\varphi_m \cos\varphi_n + \sin\varphi_m \sin\varphi_n \tag{I.42}$$

$$\begin{cases}\cos\varphi_m = \dfrac{R}{Z_m} \\ \cos\varphi_n = \dfrac{R}{Z_n}\end{cases} \quad \text{and} \quad \begin{cases}\sin\varphi_m = \dfrac{m\omega L}{Z_m} \\ \sin\varphi_n = \dfrac{n\omega L}{Z_n}\end{cases} \tag{I.43}$$

$$\begin{cases}U_m = I_m Z_m \\ U_n = I_n Z_n\end{cases} \tag{I.44}$$

We can rewrite (I.41):

$$D^2 = \sum_{m=1}^{\infty} \sum_{\substack{n=1 \\ n \neq m}}^{\infty} [(m-n)\omega L I_m I_n]^2 \tag{I.45}$$

We calculate the "field mass" using (I.40):

$$m_{field}(t) = \frac{L}{c_0^2} \sum_{m=1}^{\infty} \sum_{\substack{n=1 \\ n \neq m}}^{\infty} \cos[(m-n)\omega t + \alpha_{m-n} - \varphi_{m-n}] - \frac{L}{c_0^2} \sum_{m=1}^{\infty} \sum_{\substack{n=1 \\ n \neq m}}^{\infty} \cos[(m+n)\omega t + \alpha_{m+n} - \varphi_{m+n}] \tag{I.46}$$

Where:
$$\alpha_{m-n} = \alpha_m - \alpha_n;$$
$$\alpha_{m+n} = \alpha_m + \alpha_n$$

And:
$$\varphi_{m-n} = \varphi_m - \varphi_n;$$
$$\varphi_{m+n} = \varphi_m + \varphi_n$$

Starting to the derivative function of time for (I.46), we obtain the inertial force:

$$F_{in}(t) = \frac{L}{c_0} \sum_{m=1}^{\infty} \sum_{\substack{n=1 \\ n \neq m}}^{\infty} \frac{I_m I_n}{2} [-(m-n)\sin[(m-n)\omega t + \alpha_{m-n} - \varphi_{m-n}] + (m+n)\omega \sin[(m+n)\omega t + \alpha_{m+n} - \varphi_{m+n}]]$$

(I.47)

Using (I.18) for each harmonic, we calculated the effective value of the inertial force:

$$F_{ef.in} = 2\left[\sum_{m=0}^{\infty} \frac{1}{T}\left[\underbrace{\int_0^{T/2m} F_{in}^2(t)dt + \ldots + \int_{(2m-1)\frac{T}{2m}}^{2m\frac{T}{2m}} F_{in}^2(t)dt}_{\times m}\right] + \sum_{\substack{n=0 \\ n \neq m}}^{\infty} \frac{1}{T}\left[\underbrace{\int_0^{T/2n} F_{in}^2(t)dt + \ldots + \int_{(2n-1)\frac{T}{2n}}^{2n\frac{T}{2n}} F_{in}^2(t)dt}_{\times n}\right]\right]^{1/2}$$

Taking under consideration the orthogonally of the trigonometric functions, we obtain:

$$F_{ef.in}^2 \cdot c_0^2 = \sum_{k=1}^{\infty} (k\omega L I_k^2)^2 + \sum_{m=1}^{\infty} \sum_{\substack{n=1 \\ n \neq m}}^{\infty} (m^2 + n^2)\omega^2 L^2 (I_m I_n)^2 \tag{I.48}$$

The above relation is the multiplication between the square power produced by the inertial force and the square of speed of light.
We calculate the sum $Q^2 + D^2$, starting with the reactive and distorting powers given by C. Budeanu:

$$Q = \sum_{k=1}^{\infty} U_k I_k \sin \varphi_k = \sum_{k=1}^{\infty} (k\omega L I_k^2) \tag{I.49}$$

$$D^2 = \sum_{m=1}^{\infty} \sum_{\substack{n=1 \\ n \neq m}}^{\infty} [(m-n)\omega L I_m I_n]^2 \tag{I.50}$$

$$Q^2 + D^2 = \sum_{k=1}^{\infty} (k\omega L I_k^2)^2 + \sum_{m=1}^{\infty} \sum_{\substack{n=1 \\ n \neq m}}^{\infty} (m^2 + n^2)(\omega L)^2 I_m^2 I_n^2 \tag{I.51}$$

We can observe that (I.48) and (I.51) are identically.
In conclusion, the square of the power given by the effective value of the inertial force is equal to the sum of the reactive and distorting powers, according C. Budeanu and S. Fruze [2], [3].

### I.3.1. The decomposition under physics criteria

We decompose relation(I.48) using electrical considerations.
Based by superposition principle, the electrical current into a circuit with non-sinusoidal periodic voltage, is equal to the sum of the currents generated by each harmonic of voltage, if only itself operation in the circuit.
Based to relation (I.27), the reactive power generated by the *k* harmonic is equal:

$$Q_k = k\omega L I_k^2$$

The reactive power generated by all harmonics is equal:

$$Q = \sum_{k=1}^{\infty} k\omega L I_k^2 \qquad (I.52)$$

Therefore, we obtain an identically expression that of C. Budeanu.
The distorting power proposes by C. Budeanu result from rel.(I.51).
The decomposing of the reactive power after S. Fruze, in two orthogonal components: reactive power and distorting power from C. Budeanu, has based on the superposition principle using in the electric circuit theory.
Reactive power after C. Budeanu represents the interaction between the current harmonics of the same order, $m = n$.
Distorting power is the interaction between the current harmonics of different order, $m \neq n$.

### I.4. The electrical apparent power-S

In an electrical circuit two forces work the Coulomb force, that is displace with very small velocity and the inertial force of the "field mass" that is displace with speed of light.
Our scope is to determine a Coulomb equivalent force that is displacing with the same velocity with the inertial force of the "field mass". Adding these two forces, we obtain a resulting force that is displacing with speed of light. His effective value determines the apparent power.
The expression that gives the equivalent electric force must have the same form that the tension applies:

$$F_{echiv}(t) = K \sin \omega t \qquad (I.53)$$

The constant *K* has determined from the following relation:

$$c_0^2 F_{ef.echiv}^2 = c_0^2 \frac{1}{T} \int_0^T K^2 \sin^2 \omega t = (UI \cos \varphi)^2$$

$$K = \frac{\sqrt{2} UI \cos \varphi}{c_0} \qquad (I.54)$$

We substitute (I.54) to (I.53) and we obtain the equivalent electric force:

$$F_{echiv}(t) = \frac{\sqrt{2}UI\cos\varphi}{c_0}\sin\omega t \tag{I.55}$$

Using rel.(I.23), (I.27), (I.28) and rel.(I.55) we obtain the resultant force from the electric circuit:

$$F_{rez}(t) = \frac{\sqrt{2}UI\cos\varphi}{c_0}\sin\omega t + \frac{UI\sin\varphi}{c_0}\sin 2\omega t \tag{I.56}$$

The two forces given in rel.(I.55) and rel.(I.23) are orthogonal under a period.
The effective value for the resultant force given to rel. (I.56) is equal:

$$F_{ef.rez} = \frac{1}{c_0}UI \tag{I.57}$$

The apparent power is equal:

$$S = F_{ef.rez}c_0 = UI \tag{I.58}$$

Under non-sinusoidal condition, rel. (I.56) becomes:

$$F_{rez}(t) = \sum_{k=0}^{\infty}\frac{\sqrt{2}U_k I_k \cos\varphi_k \sin k\omega t}{c_0} - \sum_{k=0}^{\infty}\frac{U_k I_k \sin\varphi_k \sin 2k\omega t}{c_0} \tag{I.59}$$

The first term represent the equivalent Coulomb force $\overline{F}_{echiv}$ and the second represent the inertial force of the "field mass":

$$F_{echiv}(t) = \sum_{k=0}^{\infty}\frac{\sqrt{2}U_k I_k \cos\varphi_k}{c_0}\sin k\omega t \tag{I.60}$$

$$F_{in}(t) = \sum_{k=0}^{\infty}\frac{U_k I_k \sin\varphi_k}{c_0}\sin 2k\omega t \tag{I.61}$$

The two forces are orthogonally, that is:

$$\langle \overline{F}_{echiv}(t), \overline{F}_{in}(t) \rangle = 0$$

The effective value to the resultant force given (I.59) is equal:

$$\left\|\overline{F}_{ef.rez}\right\|^2 = \left\|\overline{F}_{echiv}\right\|^2 + \left\|\overline{F}_{in}\right\|^2 \tag{I.62}$$

The effective value of the Coulomb force equivalent is equal:

$$\left\|\overline{F}_{echiv}\right\|^2 = \frac{1}{T}\int_0^T\left(\sum_{k=0}^{\infty}\frac{\sqrt{2}U_k I_k \cos\varphi_k}{c_0}\sin k\omega t\right)^2 dt = \frac{1}{c_0^2}\left(\sum_{k=0}^{\infty}U_k I_k \cos\varphi_k\right)^2 = \frac{P^2}{c_0^2} \tag{I.63}$$

We substitute rel.(I.49) and rel. (I.63) into rel.(I.62), that becomes:

$$c_0^2 \|F_{rez}\|^2 = P^2 + Q^2 + S^2 = U_{ef}^2 I_{ef}^2$$

Where:

$$U_{ef}^2 = \sum_{k=0}^{\infty} U_k^2$$

and

$$I_{ef}^2 = \sum_{k=0}^{\infty} I_k^2$$

**II. The physical theory of the electrical powers in three-phased electrical circuits in sinusoidal and non-sinusoidal periodic regime**

**II.1. The calculus of the electrical powers corresponding to inertial forces in three-phased electrical circuits**

**II.1.1. Unbalanced three-phased electrical circuits in sinusoidal periodic regime**

We consider a three-phased electrical network, which feeds a three-phased consumer with the impedances:

$$Z_1 = (R_1, \omega L_1), \ Z_2 = (R_2, \omega L_2), \ Z_3 = (R_3, \omega L_3)$$

The electrical currents $i_1(t)$, $i_2(t)$, $i_3(t)$ which pass through the tree impedances are:

$$i_i(t) = \sqrt{2} I_i \sin(\omega t + \alpha_i - \varphi_i) \qquad i = 1,2,3 \qquad (II.1)$$

The instant magnetic energy on each phase is:

$$W_{m,i}(t) = \frac{L_i i_i^2(t)}{2} = L_i I_i^2 \sin^2(\omega t + \alpha_i - \varphi_i) \qquad i = 1,2,3 \qquad (II.2)$$

According to axiom A1), the field mass corresponding to relations (II.2), on each phase, is:

$$m_{field,i}(t) = \frac{L_i I_i^2 \sin^2(\omega t + \alpha_i - \varphi_i)}{c_0^2} \qquad i = 1,2,3 \qquad (II.3)$$

The magnetic energy (II.2), even though is variable in time and propagates with the velocity of light $c_0$, is not radiated, because it depends on the velocity of electrons [4].
The impulse associated to the field mass, has the form: $p = c_0 m_{field}$

$$p_i(t) = \frac{L_i I_i^2 \sin^2(\omega t + \alpha_i - \varphi_i)}{c_0} \qquad i = 1,2,3 \qquad (II.4)$$

According to axiom A2), the inertial force, corresponding to relations (II.4), is:

$$F_{in}(t)_i = \frac{dp_i}{dt} = \frac{\omega L_i I_i^2 \sin 2(\omega t + \alpha_i - \varphi_i)}{c_0} \qquad i = 1,2,3 \qquad (II.5)$$

We calculate the symmetrical components of the forces from relation (II.5) using the complex representation:

$$\begin{cases} \underline{F}_{in.1} = F_{1x} \\ \underline{F}_{in.2} = F_{2x} + jF_{2y} \\ \underline{F}_{in.3} = F_{3x} + jF_{3y} \end{cases} \quad \text{where } j = \sqrt{-1} \qquad (II.6)$$

The symmetrical components of the forces from relation (II.6) have the expression:

$$\begin{cases} \underline{F}_h = \frac{1}{3}(\underline{F}_{in.1} + \underline{F}_{in.2} + \underline{F}_{in.3}) = \frac{1}{3}\left[F_{1x} + F_{2x} + F_{3x} + j(F_{2y} + F_{3y})\right] \\ \underline{F}_d = \frac{1}{3}(\underline{F}_{in.1} + a\underline{F}_{in.2} + a^2\underline{F}_{in.3}) = \frac{1}{3}\left[\left(F_{1x} - \frac{1}{2}F_{2x} - \frac{\sqrt{3}}{2}F_{2y} - \frac{1}{2}F_{3x} + \frac{\sqrt{3}}{2}F_{3y}\right) + j\left(\frac{\sqrt{3}}{2}F_{2x} - \frac{1}{2}F_{2y} - \frac{\sqrt{3}}{2}F_{3x} - \frac{1}{2}F_{3y}\right)\right] \\ \underline{F}_i = \frac{1}{3}(\underline{F}_{in.1} + a^2\underline{F}_{in.2} + a\underline{F}_{in.3}) = \frac{1}{3}\left[\left(F_{1x} - \frac{1}{2}F_{2x} + \frac{\sqrt{3}}{2}F_{2y} - \frac{1}{2}F_{3x} - \frac{\sqrt{3}}{2}F_{3y}\right) + j\left(-\frac{\sqrt{3}}{2}F_{2x} - \frac{1}{2}F_{2y} + \frac{\sqrt{3}}{2}F_{3x} - \frac{1}{2}F_{3y}\right)\right] \end{cases}$$
(II.7)

The total initial force on the three phases is obtained from relation (II.7) and has the following expression:

$$F_{in,total}^2 = 3\left(\left|\underline{F}_h\right|^2 + \left|\underline{F}_d\right|^2 + \left|\underline{F}_i\right|^2\right) = F_{in,1}^2 + F_{in,2}^2 + F_{in,3}^2 \qquad (II.8)$$

The power that corresponds to the inertial force constant in time, $F_{in,ef}$ which is moving with the speed of light, is calculated from (II.8)

$$c_0^2 F_{in,ef}^2 = \left(\sum_{i=1}^{3} \omega L_i I_i^2\right)^2 + \sum_{\substack{i=1 \\ i \neq j}}^{3} \sum_{j=1}^{3} \omega^2 (L_i - L_j)^2 I_i^2 I_j^2 \qquad (II.9)$$

There are the following relations in function of the phases' parameters:

$$\cos\varphi_i = \frac{R_i}{Z_i}; \cos\varphi_j = \frac{R_j}{Z_j}; \sin\varphi_i = \frac{\omega L_i}{Z_i}; \sin\varphi_j = \frac{\omega L_j}{Z_j}$$

Which we introduce in (II.9), leads to:

$$c_0^2 F_{in,ef}^2 = \left(\sum_{i=1}^{3} U_i I_i \sin\varphi_i\right)^2 + \sum_{\substack{i=1 \\ j \neq i}}^{3} \sum_{j=1}^{3} \left[U_i^2 I_j^2 + U_j^2 I_i^2 - 2U_i U_j I_i I_j \cos(\varphi_i - \varphi_j)\right] \qquad (II.10)$$

From relation (II.10), one can observe that in an unbalanced electrical circuit, the inertial force of the field mass leads to the appearance of two electrical powers in the electrical circuit: the reactive power [1], the asymmetrical reactive power [6]:

$$\begin{cases} Q = \sum_{i=1}^{3} U_i I_i \sin\varphi_i \\ Q_{asy}^2 = \sum_{i=1}^{3} \sum_{\substack{j=1 \\ j \neq i}}^{3} \left[ U_i^2 I_j^2 + U_j^2 I_i^2 - 2U_i U_j I_i I_j \cos(\varphi_i - \varphi_j) \right] \end{cases} \quad (\text{II.11})$$

### II.1.2. Unbalanced three-phased electrical circuits in non-sinusoidal periodic regime

Let us consider a three-phased electrical network, with the parameters on phases $(R_1,L_1)$, $(R_2,L_2)$, $(R_3,L_3)$, in which there are the periodic, non-sinusoidal electrical currents:

$$i_i(t) = \sum_{\mu=1}^{\infty} \sqrt{2} I_{i\mu} \sin(\mu\omega t + \alpha_{i\mu} - \varphi_{i\mu}) \qquad i = 1,2,3 \quad (\text{II.12})$$

We follow the same reasoning as the one presented in paragraph II.1.1 and we obtain the generalization of the relation (II.10) for non-sinusoidal periodic regime:

$$c_0^2 F_{in,ef}^2 = \left[ 3 \sum_{i=1}^{3} \left( \sum_{\alpha=1}^{\infty} U_{i\alpha} I_{i\alpha} \sin\varphi_{i\alpha} \right) \right]^2 + \sum_{i=1}^{3} \sum_{\substack{j=1 \\ j \neq i}}^{3} \left[ \sum_{\mu=1}^{\infty} \sum_{\nu=1}^{\infty} U_{i\mu}^2 I_{j\nu}^2 + U_{j\nu}^2 I_{i\mu}^2 - 2U_{i\mu} U_{j\nu} I_{i\mu} I_{j\nu} \cos(\varphi_{i\nu} - \varphi_{j\nu}) \right] \quad (\text{II.13})$$

From relation (II.8), one can observe that in an unbalanced electrical circuit, in non-sinusoidal periodic regime, the inertial force of the field mass leads to the appearance of four electrical powers in the electrical circuit:
- The reactive power:

$$Q = \sum_{i=1}^{3} \left( \sum_{\alpha=1}^{\infty} U_{i\alpha} I_{i\alpha} \sin\varphi_{i\alpha} \right) \quad (\text{II.14})$$

which is the sum of the reactive powers corresponding to each harmonica, summed on all the phases.
- The asymmetrical reactive power on harmonics :

$$Q_{asy}^2 = \sum_{i=1}^{3} \sum_{\substack{j=1 \\ j \neq i}}^{3} \left[ \sum_{\mu=1}^{\infty} U_{i\mu}^2 I_{j\mu}^2 + U_{j\mu}^2 I_{i\mu}^2 - 2U_{i\mu} U_{j\mu} I_{i\mu} I_{j\mu} \cos(\varphi_{i\mu} - \varphi_{j\mu}) \right] \quad (\text{II.15})$$

which represents the interaction between two harmonics of the same order between two different phases. This situation means that we can divide the three-phased system in a sum of three-phased electrical systems in which activates one by one each harmonica. Each of the component circuits will have an asymmetry power.
- the deforming power:

$$D^2 = \sum_{i=1}^{3} \left[ \sum_{\mu=1}^{\infty} \sum_{\substack{\nu=1 \\ \nu \neq \mu}}^{\infty} U_{i\mu}^2 I_{i\nu}^2 + U_{i\nu}^2 I_{i\mu}^2 - 2U_{i\mu} U_{i\nu} I_{i\mu} I_{i\nu} \cos(\varphi_{i\mu} \cdot \varphi_{i\nu}) \right] \quad (\text{II.16})$$

which represents the interaction between different harmonics of the same phase.
- The deforming power of asymmetry:

$$D_{asy}^2 = \sum_{i=1}^{3} \sum_{\substack{j=1 \\ j \neq i}}^{3} \left[ \sum_{\mu=1}^{\infty} \sum_{\substack{\nu=1 \\ \nu \neq \mu}}^{\infty} U_{i\mu}^2 I_{j\nu}^2 + U_{j\nu}^2 I_{i\mu}^2 - 2U_{i\mu} U_{j\nu} I_{i\mu} I_{j\nu} \cos(\varphi_{i\mu} - \varphi_{j\nu}) \right] \quad (\text{II.17})$$

The deforming power of asymmetry represents the interaction between two different harmonics situated on different phases and it also represents a generalization of the deforming power, introduced by C. Budeanu in mono-phased electrical circuits.

### II.2. The active electrical power

In three-phased electrical circuits, the calculus of the active electrical power has as a starting point the product between the active force of electrical nature, which activates on electrical charges and their velocity.

Let $\rho_i, i = 1,2,3$ be the density of electrical charges in the three phases of the consumer. Their velocity is:

$$v_i(t) = \sqrt{2} V_i \sin(\omega t + \alpha_i - \varphi_i) \qquad i = 1,2,3 \qquad (II.18)$$

in the electrical filed of the resistive consumer.

$$E_{R,i}(t) = \sqrt{2} E_i \cos\varphi_i \sin(\omega t + \alpha_i) \qquad i = 1,2,3 \qquad (II.19)$$

The active electrical forces on phase that operates on the electrical charges are:

$$F_i(t) = \rho_i E_{R,i} = \rho_i \sqrt{2} E_i \cos\varphi_i \sin(\omega t + \alpha_i) \qquad i = 1,2,3 \qquad (II.20)$$

The instantaneous electrical power on each phase can be obtained from (II.18) and (II.20):

$$p_i(t) = v_i(t) F_i(t) = 2 V_i E_i \rho_i \cos\varphi_i \sin(\omega t + \alpha_i)\sin(\omega t + \alpha_i - \varphi_i) \qquad (II.21)$$

Based on Ohm's law in local form:

$$J_i = \rho_i V_i = \sigma E_i \qquad i = 1,2,3 \qquad (II.22)$$

Relation (II.20) becomes:

$$p_i(t) = 2\sigma E_i^2 \cos\varphi_i \sin(\omega t + \alpha_i)\sin(\omega t + \alpha_i - \varphi_i) \qquad (II.23)$$

The electrical energy on a period will be equalled to the electrical energy of an electrical source constant in time:

$$2\sigma E_i^2 \cos\varphi_i \int_0^T \sin(\omega t + \alpha_i)\sin(\omega t + \alpha_i - \varphi_i) dt = \sigma E_{c,i}^2 T \qquad i = 1,2,3 \qquad (II.24)$$

By direct calculus, we obtain the expression of the electrical field, constant in time, $E_{c,i}$ of the direct current sources:

$$E_{c,i} = E_i \cos\varphi_i \qquad i = 1,2,3 \qquad (II.25)$$

The electrical field of the equivalent sources operates with a constant force in time on the density of the electrical charges $\rho_i$, which move with a constant velocity $v_i$. In the end, there is obtained the density of power in direct current.

$$p_c = \rho_i V_i \cos\varphi_i E_i = J_i E_i \cos\varphi_i \qquad i = 1,2,3 \qquad (II.26)$$

By integrating on the whole volume, we obtain the total power:

$$P_{total} = \sum_{i=1}^{3} U_i I_i \cos\varphi_i \qquad (II.27)$$

**II.3. Apparent electrical power**

In an electrical consumer, there are two types of forces: active forces of electrical nature and inertial forces. The active electrical forces move with an velocity $v_i$, which is variable in time, meanwhile the inertial forces move with the velocity of light $c_0$, which is constant In time. In order to compose these two forces, it is necessary to calculate the equivalent of the active electrical forces, which have to be variable in time and move with the velocity of light $c_0$. There is shown that it has the expression:

$$F_{echiv}(t)_i = \frac{\sqrt{2}U_i I_i \cos\varphi_i}{c_0}\sin(\omega t + \alpha_i) \qquad i = 1,2,3 \qquad (II.28)$$

And the inertial force on each phase:

$$F_{in,i}(t) = \frac{\sqrt{2}U_i I_i \sin\varphi_i}{c_0}\sin(2\omega t + 2\alpha - 2\varphi_i) \qquad i = 1,2,3 \qquad (II.29)$$

The resulting force on each phase is obtained from (II.28) and (II.29):

$$F_{rez,i}(t) = \frac{\sqrt{2}U_i I_i \cos\varphi_i}{c_0}\sin(\omega t + \alpha_i) + \frac{\sqrt{2}U_i I_i \sin\varphi_i}{c_0}\sin(2\omega t + 2\alpha_i - 2\varphi_i) \qquad (II.30)$$

The total force on all the phases is:

$$F_{total}(t) = \sum_{i=1}^{3}\left[\frac{\sqrt{2}U_i I_i \cos\varphi_i}{c_0}\sin(\omega t + \alpha_i) + \frac{\sqrt{2}U_i I_i \sin\varphi_i}{c_0}\sin 2(\omega t + \alpha_i - 2\varphi_i)\right] \qquad (II.31)$$

The effective value of the total force is:

$$F_{total}^2 = \frac{1}{c_0^2}(U_1^2 + U_2^2 + U_3^2)(I_1^2 + I_2^2 + I_3^2) \qquad (II.32)$$

The expression:

$$c_0^2 F_{total}^2 = (U_1^2 + U_2^2 + U_3^2)(I_1^2 + I_2^2 + I_3^2) = S_{total}^2 \qquad (II.33)$$

represents the total apparent electrical power in three-phased electrical circuits.
For three-phased electrical circuits, in non-sinusoidal periodic regime, relations (II.28) and (II.29) become:

$$F_{echiv,asy}(t) = \sum_{\mu=1}^{\infty} \frac{\sqrt{2} U_{\mu i} I_{\mu i} \cos \varphi_{\mu i}}{c_0} \sin(\mu \omega t + \alpha_{\mu i}) \qquad i = 1,2,3 \qquad \text{(II.34)}$$

$$F_{in,asy,i}(t) = \sum_{\mu=1}^{\infty} \frac{\sqrt{2} U_{\mu i} I_{\mu i} \sin \varphi_{\mu i}}{c_0} \sin(2\mu \omega t + 2\alpha_{\mu i} - 2\varphi_i) \qquad i = 1,2,3 \qquad \text{(II.35)}$$

The total force on the three phases can be obtained from (II.33) and (II.34).

$$F_{total,asy}(t) = \sum_{i=1}^{3} \left| F_{echiv,asy,i}(t) + F_{in,asy,i}(t) \right| \qquad \text{(II.36)}$$

The effective value of relation (II.35) is:

$$c_0^2 F_{total,asy}^2 = \left( \sum_{i=1}^{3} \sum_{\mu=1}^{\infty} U_{i\mu}^2 \right) \left( \sum_{i=1}^{3} \sum_{\mu=1}^{\infty} I_{i\mu}^2 \right) \qquad \text{(II.37)}$$

and it represents the total electrical power in three-phased electrical circuits, in non-sinusoidal periodic regime. From the analysis of relations (II.33) and (II.37), it results that a continuous source of electrical energy has the voltage and current equal to:

$$U_{echiv} = \sqrt{U_1^2 + U_2^2 + U_3^2} \qquad I_{echiv} = \sqrt{I_1^2 + I_2^2 + I_3^2} \qquad i = 1,2,3 \qquad \text{(II.38)}$$

For non-sinusoidal periodic regime, the effective voltages on phase are:

$$U_i = \sqrt{U_{i1}^2 + U_{i2}^2 + .......... + U_{ik}^2 + ....} \qquad i = 1,2,3 \qquad k = 1,2....\infty \qquad \text{(II.39)}$$

And the effective currents are:

$$I_i = \sqrt{I_{i1}^2 + I_{i2}^2 + .......... + I_{ik}^2 + ....} \qquad i = 1,2,3 \qquad k = 1,2....\infty \qquad \text{(II.40)}$$

From relations (II.14), (II.15), (II.16), (II.17), (II.27) and (II.37), it results that for three-phased electrical circuits in non-sinusoidal periodic regime:

$$P^2 + Q^2 + Q_{asy}^2 + D^2 + D_{asy}^2 = S^2 \qquad \text{(II.41)}$$

### III. Conclusions

1. The physical essence of power components is determined as the main aim of the article. This problem we been solving using only the effective values for the Coulomb forces and inertial forces over a period.

2. The equivalent effective Coulomb forces produce a mechanic work that is the electric active power.
3. The effective inertial forces of the "field mass" and superposition principle produce a mechanical work that is reactive and distorting powers.
4. The effective resultant force that is obtained from equivalent Coulomb forces an inertial forces of the "field mass' produce a mechanical work that is apparent power.
5. The inertial forces lead to: the reactive three-phased electrical power Q, the asymmetrical three-phased electrical power - $P_{asy}$, the deforming three-phased electrical power D, the deforming three-phased power of asymmetry - $D_{asy}$.
6. The composing of the active electrical forces with the inertial electrical forces lead to the apparent three-phased power S.
7. Between the six three-phased electrical powers, there is the following relation:

$$P^2 + Q^2 + Q_{asy}^2 + D^2 + D_{asy}^2 = S^2$$

The all results may synthesize in a diagram, which presents the logical relationships between the fundamental elements of the theory presented and the consequences immediate.

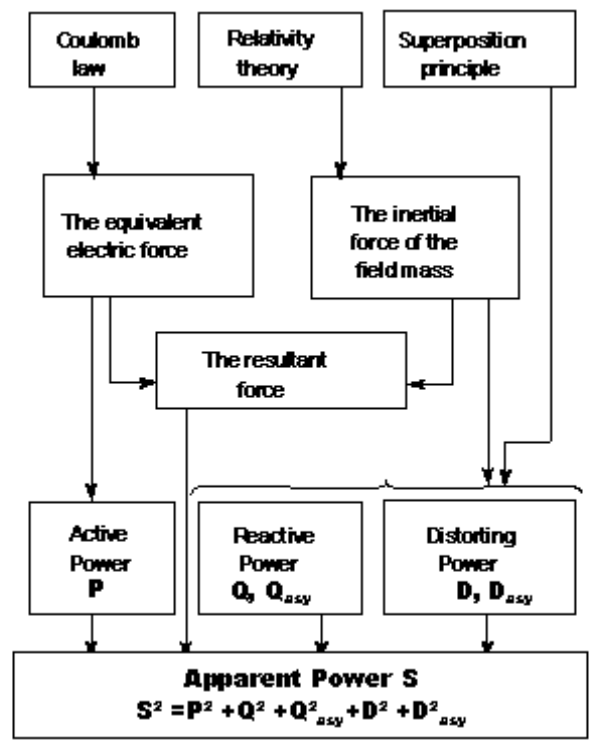

# References

[1] C.I. Budeanu, *Puissance réactives et fictives,* Editura I.P.E., Bucuresti, 1927.

[2] M. Slonim, *Physical essence of power components,* L' Energia Elettrica, vol.81, 2004, Ricerche.

[3] S. Sverson, *Power measurement technique for non-sinusoidal conditions,* Doctoral Thesis, Department of Electric Power Engineering Chalmers , University of Technology Göteborg Sweden, 1999.

[4] L.Landau, E. Lifchitz, *Théorie des champs,* Éditions Mir Moscou, 1970.

[5] F. Reza, *Linear Spares in Engineering,* Editura Didactica si Pedagogica, Bucuresti, 1973.

[6] Pl.Andronescu, *Fundaments of Electrotechniques*, *Vol.II*, Editura Didactica şi Pedagogica, Bucureşti, 1972.